\documentclass[twocolumn]{aastex62}

\graphicspath{{./}{figures/}}

\usepackage{color}

\received{}
\revised{}
\accepted{}

\submitjournal{ApJ}

\shorttitle{}
\shortauthors{}

\usepackage{epstopdf}
\usepackage{color}
\usepackage{graphicx}
\usepackage{longtable}

\begin{document}

\title{\large Flare Induced Sunquake Signatures in the Ultraviolet as\\ Observed by the Atmospheric Imaging Assembly}

\author{Sean Quinn}
\affil{Astrophysics Research Centre, School of Mathematics and Physics, Queen's University Belfast, BT7 1NN, Northern Ireland, UK \\
}

\author{Mihalis Mathioudakis}
\affil{Astrophysics Research Centre, School of Mathematics and Physics, Queen's University Belfast, BT7 1NN, Northern Ireland, UK \\
}

\author{Christopher J. Nelson}
\affil{Astrophysics Research Centre, School of Mathematics and Physics, Queen's University Belfast, BT7 1NN, Northern Ireland, UK \\
}

\author{Ryan O. Milligan}
\affil{Astrophysics Research Centre, School of Mathematics and Physics, Queen's University Belfast, BT7 1NN, Northern Ireland, UK \\
}

\author{Aaron Reid}
\affil{Astrophysics Research Centre, School of Mathematics and Physics, Queen's University Belfast, BT7 1NN, Northern Ireland, UK \\
}

\author{David B. Jess}
\affil{Astrophysics Research Centre, School of Mathematics and Physics, Queen's University Belfast, BT7 1NN, Northern Ireland, UK \\
}

\begin{abstract}

Sunquakes (SQs) have been routinely observed in the solar photosphere, but it is only recently that signatures of these events have been detected in the chromosphere. We investigate whether signatures of SQs are common in Ultraviolet (UV) continua, which sample the solar plasma several hundred km above where SQs are typically detected. We analyse observations from the Solar Dynamics Observatory's Atmospheric Imaging Assembly (SDO/AIA) 1600 \AA\ and 1700 \AA\ passbands, for SQ signatures induced by the flares of Solar Cycle 24. We base our analysis on the 62 SQs detected in the recent statistical study presented by \citet{Sharykin2020}. We find that 9 out of 62 SQ candidates produced a response that is clearly detected in running difference images from the AIA 1600 \AA\ and 1700 \AA\ channels. A binary frequency filter with a width of $2$ mHz, centred on $6$ mHz, was applied to the data. The first signature of each SQ was detected at distances between 5.2 Mm to 25.7 Mm from the associated flare ribbon. Time-distance and regression analysis allowed us to calculate the apparent transverse velocities of the SQs in the UV datasets and found maximum velocities as high as 41 km s$^{-1}$, 87 Mm away from the SQ source. Our analysis shows that flare induced SQ signatures can be detected in the SDO/AIA 1600 \AA\ and 1700 \AA\ passbands, hinting at their presence in the lower chromosphere. There was no apparent correlation between GOES flare classification, and the appearance of the SQ at these heights.  

\end{abstract}

\keywords{Sun: chromosphere, Sun: quakes, Sun: flares, Sun: helioseismology}

\section{Introduction} 
\label{sec:intro}

The photospheric perturbations which often accompany solar flares can generate helioseismic waves, commonly known as ``sunquakes" (SQs). The suggestion that solar flares could produce some form of seismic response in the solar interior was first reported by \citep{Wolff} and more recently investigated by \citep{KosovichvZharkova1995}. The first detection of a SQ created by a solar flare was detected by \cite{KosovichevZharkova1998}, using the Michelson Doppler Interferometer (MDI; \citealt{Scherrer1995}) onboard the Solar and Heliospheric Observatory (SOHO; \citealt{Domingo1995}). SQs manifest as expanding, anisotropic waves in running difference dopplergrams.   

The creation of SQs is usually attributed to dense plasma refracting acoustic wavefronts which are travelling into the solar interior. It is thought that hydrodynamic shocks are created when a beam of high-energy particles, either an electron beam \citep{KosovichevZharkova1998, Kosovichev2007} or a proton beam \citep{ZharkovaZharkov2007}, is accelerated deep into the solar atmosphere from the corona during a flare, depositing large amounts of energy and momentum via collisions with the atmospheric plasma. The collisions result in heating and the generation of EUV and soft X-Ray emission through the process of chromospheric evaporation. The resulting overpressure leads to chromospheric condensation pushing the plasma deeper into the solar atmosphere, where it collides with the denser photosphere, or dissipates after some time. The particle beams create a high pressure compression in the photosphere, and the generation of an acoustic shock front propagating into the internal layers of the Sun \citep{Kosovichev2006}. The acoustic shock is reflected due to temperature and density fluctuations in the solar interior, causing a change in the sound speed \citep{Green2017} which results in the creation of expanding ripples on the solar surface. This theory of SQ generation could be linked to the thick-target model, and arguments have been made for SQ generation to be added to these models \citep{Kosovichev2014, Sharykin2018}.

\begin{figure*}
\includegraphics[width=\textwidth]{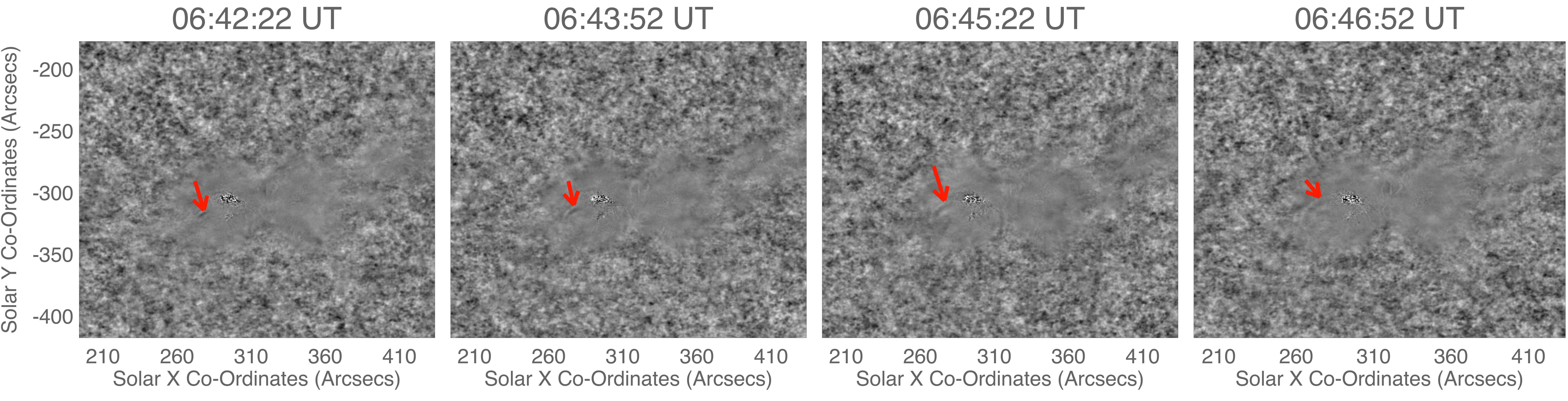}
\caption{A 240\arcsec$\times$240\arcsec\ field-of-view (FOV) image constructed using a running difference (specifically, fr[n+1]-fr[n]) technique on the SDO/HMI LOS for the M5.2 flare from 2014 February 02, plotted at four time-steps. The time of each panel is indicated in the title of each panel. The red arrows point to one of the propagating wavefronts of the SQ, but more can be seen in the bottom right of the flare ribbon.}
\label{Fig1}
\end{figure*}

The acceptance that SQs are an intrinsic aspect of a model such as the thick-target model would imply SQs should be present in most solar flares. However, SQs are still observed events during a fraction of the solar flares that occur on the Sun. With this in mind, there are a number of alternative interpretations of SQ generation that have been proposed in the literature. Potential scenarios for SQ excitation include: Heating by continuum radiation \citep{lindsey_and_donnea2008}; Wave heating of the plasma \citep{Matthews2015}; And magnetic field changes and the creation of a Lorentz force \citep{Hudson2008, Fisher2012}. Interestingly, \cite{Zharkov2011} reported SQs whose acoustic sources were detected away from their hard X-Ray sources, indicating that the SQs were not formed at the location of particle injection. They concluded that the magnetic field reconfiguration was responsible for a flux rope eruption and the subsequent creation of the SQ. The launch of the Solar Dynamics Observatory (SDO; \citealt{Pesnell}), which carries the Helioseismic and Magnetic Imager (HMI; \citealt{Scherrer12}), has allowed the wide-spread detection of flare-induced SQs over the past decade \citep{Kosovichev2014}.

To date, the majority of detections of the seismic responses associated with flares have been confined to the solar photosphere. However, high spatial and temporal resolution observations of the 2017 September 6, X9.3 flare, the largest of Solar Cycle 24, displayed evidence that chromospheric response to SQs were also possible \citep{Quinn2019}. The photospheric components of this SQ have been widely reported (\citealt{Sharykin2018, Zhao2018}). The detection of the SQ response in the chromosphere was based on observations obtained with the Swedish 1-m Solar Telescope (SST; \citealt{Scharmer2003}) Ca II 8542 \AA\ and H$\alpha$ 6563 \AA\ lines, as well as the Atmospheric Imaging Assembly (AIA; \citealt{Lemen2012}) onboard SDO. Specifically, the 1600 \AA\ and AIA 1700 \AA\ filters were studied in the work of \cite{Quinn2019}.

It is currently unknown why some flares create SQs and others do not, as well as why some SQs have responses in the chromosphere, and others only photospheric. The conversion of flare energy into acoustic waves is still a matter of great interest, as is a unified model of SQ generation. The main motivation behind this research is to investigate how common are the signatures of SQs in the SDO/AIA $1600$ \AA\ and $1700$ \AA\ continua, and if there is any correlation between the presence of any such response and the strength of the flare that created it. The starting point for this investigation is the statistical study of the SQs of Solar Cycle 24, recently presented by \cite{Sharykin2020}.  A sub-set of events from this sample are investigated to determine if these SQs produce a detectable signature in the SDO/AIA 1600 \AA\ and 1700 \AA\ passbands. In Sect.~\ref{sec:Observations} we provide an overview of the observations analysed in this article; In Sect.~\ref{sec:Methodology} we present the methodology used in this analysis; In Sect.~\ref{sec:Results} we present our results; In Sect.~\ref{sec:Conclusions} we draw our conclusions.

\begin{figure*}
\includegraphics[width=\textwidth]{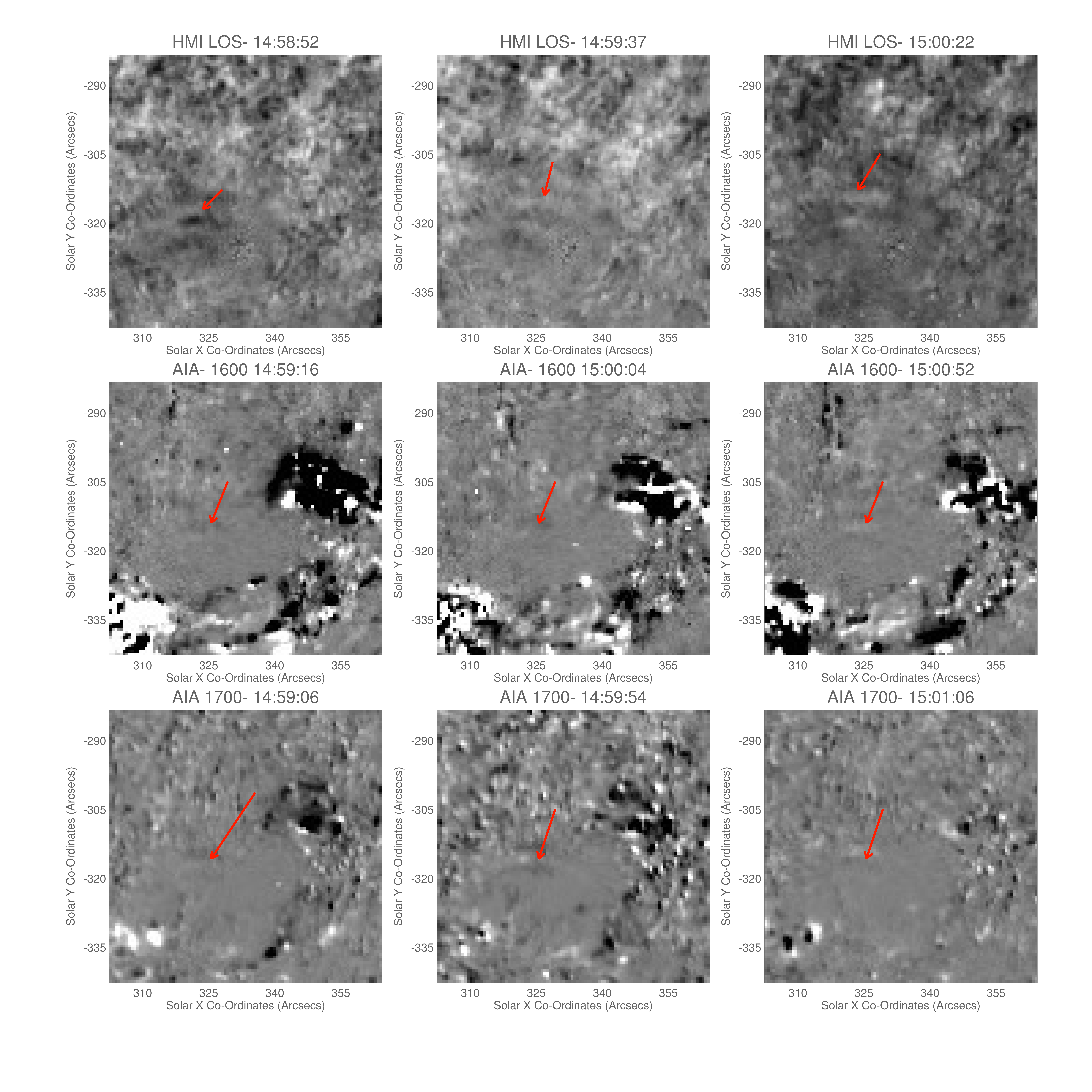}
\caption{A 60\arcsec$\times$60\arcsec\ FOV covering the M1.3 flare from 2012 February 04 including: SDO/HMI LOS running difference images (top row); SDO/AIA 1600 \AA\ running difference images (middle row); and SDO/AIA 1700 \AA\ running difference images (bottom row). Images from each channel are plotted at 3 time-steps which are as co-temporal as possible. The red arrows in each image indicate the wavefront of the SQ in the SDO/AIA channels, as it evolves across the solar surface. These images are centred on X=332\arcsec, Y=-313\arcsec. We have zoomed in from the original 240\arcsec$\times$240\arcsec\ dataset to make the wavefronts clearer.}
\label{Fig2}
\end{figure*}

\section{Observations} 
\label{sec:Observations}

The observations analysed in this article were acquired by the SDO/HMI and SDO/AIA instruments over the course of the past decade. The SDO/HMI instrument acquires both full disk line-of-sight (LOS) magnetic field, velocity, and photospheric continuum images each with a cadence of $45$ s. The SDO/HMI LOS velocity channel has been shown to best display the presence of a SQ in running difference images, and for this reason it is considered to be the photospheric representation of a SQ in this article. The signatures of SQs can be better observed by applying frequency filtering to the data. A filter centred at 6 mHz is typically used  \citep{DoneaLindsey2005, Matthews2015}. An example of a SQ in unfiltered running difference images (constructed by removing the previous frame from each image) is shown in Fig~\ref{Fig1}. This example is from the M5.2 flare of 2014 February 02 at 03:57 UT, where the SQ ripple is highlighted by the red arrow. It should be noted that running differences of the SDO/HMI continuum can also show the presence of seismic responses, however, SDO/HMI LOS velocities provide the clearest signatures. 

The SDO/AIA instrument observes a wide range of wavelengths, sampling from the photosphere to the corona, but for the purpose of this work we only investigate the emission in the Ultraviolet (UV) 1600 \AA\ and 1700 \AA\ continuum filters. The focus on these channels is based on the work of \cite{Quinn2019} who reported that SQ responses were present in data from these SDO/AIA filters for the single flare studied in that article. To build on the earlier work of \cite{Quinn2019}, we utilise the homogeneous flare list presented in \cite{Sharykin2020} which provides a comprehensive catalog of 507 M and X-class flares of Solar Cycle 24, to search for SQ signatures in the 1600 \AA\ and 1700 \AA\ channels. The full-disk data from the SDO/HMI LOS, SDO/AIA 1600 \AA, and SDO/AIA 1700 \AA\ filters were downloaded as level 1.5 data. For all channels, 1 hour and 30 minutes of data were downloaded for all relevant flares (detailed in the Sect.~\ref{sec:Methodology}), starting approximately 10 minutes before the flare peak. The SDO/HMI LOS data have a cadence of 45 s and a post-reduction pixel scale of $0.6$\arcsec, whereas these SDO/AIA UV data had a cadence of 24 s and a pixel scale of $0.6$\arcsec. The observations were cropped to a 240\arcsec$\times$240\arcsec\ FOV, centred on the active region (AR) where the flare occurred. 

In the quiet Sun, the main contribution of the intensity in the SDO/AIA 1700 \AA\ channel comes from the photospheric continuum, while the intensity in the SDO/AIA 1600 \AA\ channel comes from a combination of C IV 1550 \AA\ and photospheric continuum \citep{Lemen2012}. The intensity in these channels, therefore, samples diagnostics that originate from the upper photosphere and transition region \citep{Lemen2012}. However, in a recent paper, \cite{Simoes2019} concluded that during a flare, both the SDO/AIA 1600 \AA\ and 1700 \AA\ passbands are dominated by spectral lines that form in the chromosphere and transition region; the C IV 1550 \AA\ doublet and Si I continua for the SDO/AIA 1600 \AA\ channel and the C I 1656 \AA\ multiplet and He II 1640 \AA\ line for the SDO/AIA 1700 \AA\ channel. It should be noted that the slit spectra of \cite{Simoes2019} lacked detailed spatial information, but included emission from the ribbon in the vicinity of the flare. We are, therefore, unable to accurately assess the height at which these filters sample during the flares and SQs studied here but we are confident that these data sample a region of the solar atmosphere at least several hundred km above the region observed by the SDO/HMI instrument.

 \begin{table*}
	\centering
	\begin{tabular}{|c|c|c|c|c|c|c|c|}
\hline
\multicolumn{1}{|p{2.2cm}}{\centering Flare Start \\ Time \\(UT)} & \multicolumn{1}{|p{0.9cm}|}
{\centering GOES \\ Class} & AR & \multicolumn{1}{|p{3.2cm}}{\centering Centre of \\ Field of View \\ (Arcsecs)} & \multicolumn{1}{|p{1.5cm}|}
{\centering Visibility by \\ Eye} & \multicolumn{1}{|p{2.1cm}|}
{\centering Visibility in \\ Time-Distance Diagrams} & \multicolumn{1}{|p{1.7cm}}{\centering HMI SQ Appearance \\ Time (UT)}& \multicolumn{1}{|p{2.5cm}|}{\centering SQ Propagation \\ Direction}\\

\hline
\hline
15.02.2011 01:44 & X2.2 & 11158 & X=197\arcsec, Y=-224\arcsec & Clear & Visible & 02:28:00 & North-North East\\
\hline
30.07.2011 02:04 & M9.3 & 11261 & X=-497\arcsec, Y=195\arcsec & Faint & *Visible & 02:24:22 & South East\\
\hline
25.09.2011 08:46 & M3.1 & 11302 & X=-666\arcsec, Y=137\arcsec & Potential & Not Visible & 08:55:52 & South East\\
\hline
04.07.2012 14:35 & M1.3 & 11515 & X=334\arcsec, Y=-313\arcsec & Clear & Visible & 12:55:07 & North West\\
\hline
$\dagger$05.07.2012 06:49 & M1.1 & 11515 & X=392\arcsec, Y=-313\arcsec & Faint & Not Visible & 07:08:49 & North\\
\hline
05.07.2012 11:39 & M6.1 & 11515 & X=496\arcsec, Y=-262\arcsec & Potential & Not Visible & 12:01:52 & North\\
\hline
06.07.2012 01:37 & M2.9 & 11515 & X=576\arcsec, Y=-323\arcsec & Potential & *Visible & 02:01:52 & South West\\
\hline
09.07.2012 23:00 & M1.1 & 11520 & X=-528\arcsec, Y=-328\arcsec & Faint & Not Visible & 23:21:22 & North West\\
\hline
23.10.2012 03:13 & X1.8 & 11598 & X=-777\arcsec, Y=-265\arcsec & Clear & Visible & 03:31:52 & West\\
\hline
17.02.2013 15:45 & M1.9 & 11675 & X=-343\arcsec, Y=305\arcsec & Potential & Not Visible & 16:02:37 & West-South West\\
\hline
$\dagger$24.10.2013 10:30 & M3.5 & 11875 & X=349\arcsec, Y=31\arcsec & Faint & Not Visible & 10:51:22 & North West\\
\hline
05.11.2013 22:07 & X3.3 & 11890 & X=-639\arcsec, Y=-231\arcsec & Faint & *Visible & 22:25:37 & West\\
\hline
07.11.2013 14:15 & X2.4 & 11890 & X=-342\arcsec, Y= -238\arcsec & Faint & Unclear & 14:42:22 & North-North East\\
\hline
$\dagger$10.11.2013 05:08 & X1.1 & 11890 & X=-238\arcsec, Y= -266\arcsec & Faint & Unclear & 05:30:22 & North-North East\\
\hline
02.02.2014 06:24 & M2.6 & 11968 & X=-297\arcsec, Y=314\arcsec & Faint & Unclear & 06:54:22 & East\\
\hline
04.02.2014 03:57 & M5.2 & 11967 & X=160\arcsec, Y=-98\arcsec & Clear & Visible & 04:16:07 & North\\
\hline
$\dagger$07.02.2014 10:25 & M1.9 & 11968 & X=695\arcsec, Y=292\arcsec & Potential & Not Visble & 10:40:07 & South East\\
\hline
$\dagger$10.03.2015 03:19 & M5.1 & 12297 & X=-615\arcsec, Y=-196\arcsec & Potential & Not Visible & 03:39:22 & South West\\
\hline
22.08.2015 21:19 & M3.5 & 12403 & X=-229\arcsec, Y=-349\arcsec & Faint & *Visible & 22:37:52 & North East\\
\hline
28.09.2015 14:53 & M7.6 & 12422 & X=406\arcsec, Y=-446\arcsec & Faint & Not Visible & 15:13:07 & North East\\
\hline
$\dagger$06.09.2017 08:57 & X2.2 & 12673 & X=527\arcsec, Y=-247\arcsec & Potential & Not Visible & 09:16:52 & West\\
\hline
06.09.2017 11:53 & X9.3 & 12673 & X=557\arcsec, Y=-265\arcsec & Clear & Visible & 12:13:52 & South West\\
\hline
07.09.2017 04:49 & M2.4 & 12673 & X=677\arcsec, Y=-235\arcsec & Faint & Not Visible & 05:20:37 & South West\\
\hline
07.09.2017 10:11 & M7.3 & 12673 & X=677\arcsec, Y=-235\arcsec & Faint & Not Visible & 10:29:48 & South East\\
\hline
07.09.2017 14:20 & X1.3 & 12673 & X=732\arcsec, Y=-232\arcsec & Potential & Unclear & 14:48:22 & South East\\
\hline
\end{tabular}
\caption{The 25 flares that produced some response associated with a SQ in both the SDO/AIA 1600 \AA\ and 1700 \AA\ running difference images once the frequency filter had been applied. The approximate time of the helioseismic responses first observation in SDO/HMI LOS running differences, and approximate propagation direction are also included. Five of these SQs are clearly visible by eye, 12 have a faint visibility and 8 are considered as potential responses. The 5 that are clearly visible had varying degrees of visibility in the time-distance diagrams. Three of the faintly visible SQ responses had obvious SQ ridges in their time-distance diagrams and are also labelled as `Visible'. One of the `Potential' responses had an obvious SQ ridge in it's time-distance diagram, and is also labelled as `Visible', bringing the total of SQs detectable in SDO/AIA data to 9. Those SQs that are more obvious in the time-distance diagrams than in the running difference movies have been marked with an `$\ast$'. Six additional SQs that only possess a possible response when filtered are marked with a `$\dagger$'.}
\label{Tab1}
\end{table*}

\section{Methodology}
\label{sec:Methodology}

\begin{figure*}
\includegraphics[width=\textwidth]{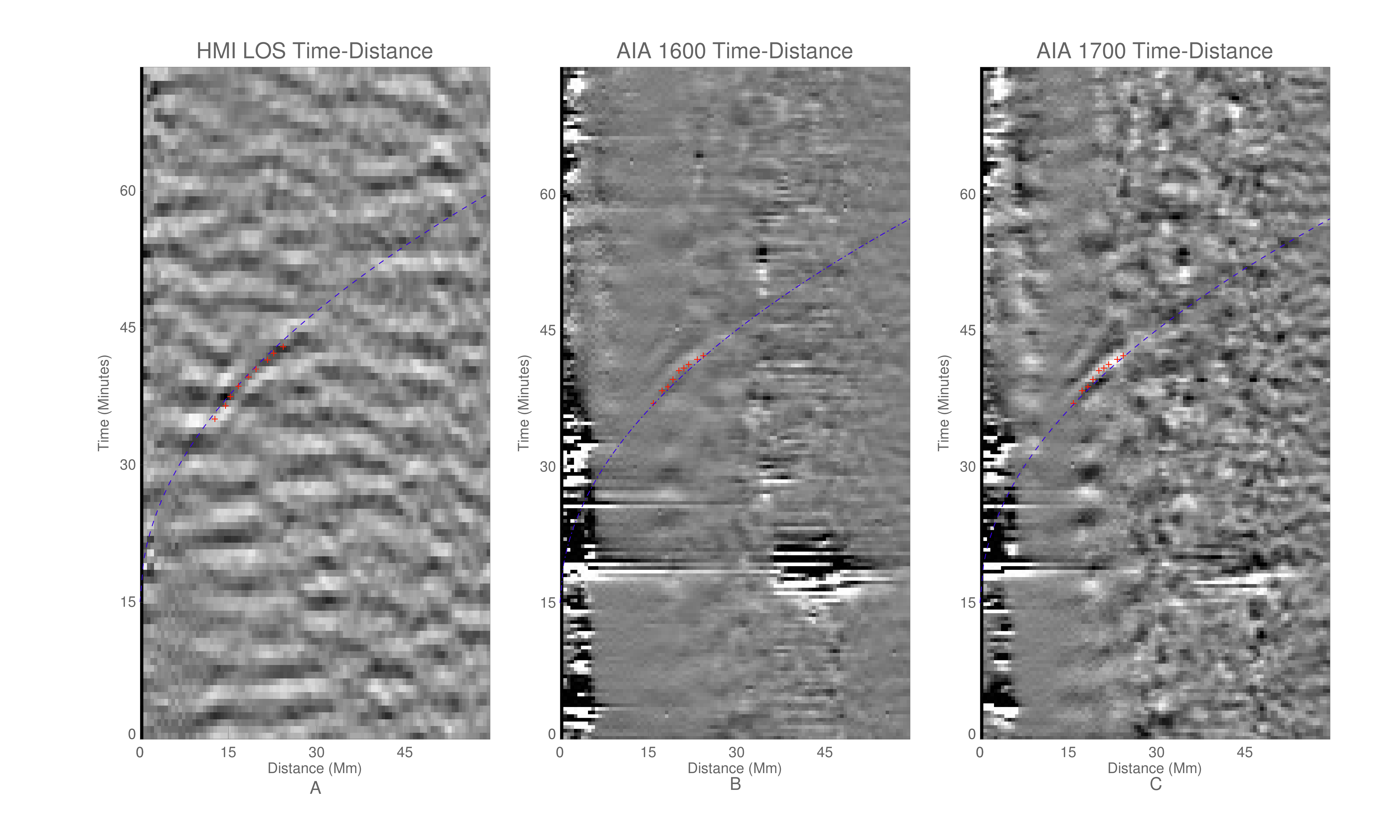}
\caption{Unfiltered time-distance diagrams for the M1.3 flare from 2012 July 04 from SDO/HMI LOS velocity (A), and temporally degraded SDO/AIA 1600 \AA\ (B) and SDO/AIA 1700 \AA\ (C) channels. The corresponding regression trends are over-plotted as blue curves. This trend has been shifted vertically so as not to obscure the ridge. The red crosses, indicate the points which were selected for the $\chi^2$ test to determine the accuracy of the fit. These points have also been offset by the same amount as the blue trend line.}
\label{Fig3}
\end{figure*}

\begin{table*}
	\centering
	\begin{tabular}{|c|c|c|c|c|c|c|c|c|c|}
\hline
\multicolumn{1}{|p{2.5cm}}{\centering Flare\\Start Time\\(UT)} & 
\multicolumn{1}{|p{0.9cm}}{\centering GOES\\Class} & 
\multicolumn{1}{|p{1.3cm}}{\centering HMI\\V$_{MAX}$\\(km s$^{-1}$)} &
\multicolumn{1}{|p{1.7cm}}{\centering HMI\\Acceleration\\(km s$^{-2}$)} &
\multicolumn{1}{|p{1.3cm}}{\centering AIA\\V$_{MAX}$\\(km s$^{-1}$)} &
\multicolumn{1}{|p{1.7cm}|}{\centering AIA\\Acceleration \\(km s$^{-2}$)}&
\multicolumn{1}{|p{1.0cm}}{\centering HMI $\chi^2$\\Value}&
\multicolumn{1}{|p{1.4cm}|}{\centering AIA $\chi^2$\\Value (Non-Degraded)}&
\multicolumn{1}{|p{1.5cm}|}{\centering AIA $\chi^2$\\Value\\(Degraded)}&
\multicolumn{1}{|p{1.6cm}|}{\centering `$\alpha$' \\Value\\HMI/AIA\\(s km$^{-\frac{1}{2}}$)}\\
\hline
\hline
$*$ 15.02.2011 01:44 & X2.2 & $37.7^{+1.5}_{-0.7}$ & 0.017 & N/A & N/A & 0.811 & 3.19 & - & 1.27/1.66 \\
\hline
30.07.2011 02:04 & M9.3 & $35.7^{+1.2}_{-0.6}$ & 0.015 & $35.7^{+0.2}_{-0.2}$ & 0.015 & 0.316 & 0.147 & 0.208 & 1.34/1.85 \\
\hline
04.07.2012 14:35 & M1.3 & $32.1^{+0.9}_{-0.4}$ & 0.012 & $32.1^{+0.2}_{-0.1}$ & 0.012& 0.69 & 0.253 & 0.566 & 1.49/1.93 \\
\hline
06.07.2012 01:37 & M2.9 & $37.0^{+1.0}_{-0.7}$ & 0.016 & $37.0^{+0.2}_{-0.2}$ & 0.016& 0.335 & 0.253 & 0.168 & 1.30/1.82 \\
\hline
$*$ 23.10.2012 03:13 & X1.8 & $36.7^{+1.1}_{-0.5}$ & 0.013 & N/A & N/A & 0.519 & 1.38 & - & 1.43/1.91 \\
\hline
05.11.2013 22:07 & X3.3 & $36.4^{+1.3}_{-0.6}$ & 0.015 & $33.5^{+0.2}_{-0.2}$ & 0.014& 0.212 & 1.87 & 0.34 & 1.32/1.86  \\
\hline
04.02.2014 03:57 & M5.2 & $41.0^{+2.0}_{-1.0}$ & 0.020 & $41.0^{+0.3}_{-0.3}$ & 0.020& 1.024 & 1.02 & 0.264 & 1.14/1.56 \\
\hline
22.08.2015 21:19 & M3.5 & $37.7^{+1.5}_{-0.7}$ & 0.017 & $33.0^{+0.2}_{-0.2}$ & 0.013& 0.508 & 0.288 & 0.249 & 1.27/1.79 \\
\hline
06.09.2017 11:53 & X9.3 & $41.0^{+1.9}_{-0.9}$ & 0.020 & $38.5^{+0.3}_{-0.3}$ & 0.017& 1.631 & 3.43 & 0.666 & 1.17/1.62 \\
\hline
\end{tabular}
\caption{The nine flares that were considered to have clear evidence of SQs in the SDO/AIA channels (from Table~\ref{Tab1}). We present the GOES Class, the apparent maximum transverse velocities and accelerations of the SQs in the SDO/HMI LOS velocity data together with the apparent maximum transverse velocities and accelerations in the SDO/AIA 1600 \AA\ and 1700 \AA\ channels. The $\chi^2$ values for the fits to the SDO/HMI LOS velocity and SDO/AIA ridges, both pre and post temporal interpolation, are also provided. The $\alpha$ value is constant in the fitted regression trends. Varying the value of `$\alpha$' in the regression trend fits (used to calculate the errors in the maximum apparent velocities) returns errors in the calculated acceleration at 3 decimal places or smaller. Therefore, no errors are reported in the acceleration columns in the table. The flares highlighted with stars displayed evidence of a SQ in their original SDO/AIA data, however, these signatures disappeared after the cadence was interpolated to match the SDO/HMI cadence. As the maximum apparent velocities and accelerations in the FOV are reported from the temporally degraded data, no velocity and acceleration values are reported for these flares. The  maximum velocities were recorded at a distance of approximately 87 Mm from the estimated epicentre of each SQ. This distance was dictated by the creation of the time-distance diagrams.}
\label{Tab2}
\end{table*}

The \citealt{Sharykin2020} investigation into the detection of flare induced SQs was based on three methods: (a) The movie method where the search for SQs was based on running difference dopplergrams; (b) The holography method which looks for acoustic sources from which a SQ would be generated; And (c) time-distance analysis where a characteristic ridge appears in a time-distance diagram. Using these detection methods, the authors determined that of the 507 flares, 181 had a photospheric perturbation, meaning the photosphere responded in some way to the flare. Of these $181$ flares, 114 resulted in the detection of at least one SQ. It is these 114 detections that form the basis for the sample studied in this article. 

The flares that displayed evidence of SQs ranged in strength from M1.0 to X9.3. We removed any SQs from our sample that had not been detected by eye (i.e. not visible in the running difference images). Those detected purely via time-distance analysis were omitted, as responses in the SDO/AIA 1600 \AA\ and 1700 \AA\ bands are extremely short lived, and for them to be picked up on a time-distance diagram it was necessary in this work to have an accurate propagation angle and distance determined by the SDO/HMI LOS response first. Finding any responses in the SDO/AIA bands using solely time-distance analysis would have been extremely difficult as the propagation angle and direction was not supplied in \cite{Sharykin2020}. We, therefore, leave this task for future work. After limiting the flares to only those present in running difference movies, the total number of events that we considered here was reduced to 62, with 49 being M-Class and 13 X-Class. These flares are listed in Table~\ref{Tab3}, in the Appendix. 
 
We applied a binary frequency filter centred at 6 mHz, with a width of $2$ mHz, to all datasets  \citep{Jess2017}. Using these filtered datasets, we began our analysis by investigating the SDO/HMI observations of the 62 flares to confirm all SQ detections. This aided our search for SQ signatures in the SDO/AIA 1600 \AA\ and 1700 \AA\ channels. For every SQ identified in the SDO/HMI LOS velocity running difference images, the SDO/AIA running differences were also investigated. Looking in the same region as the wavefronts in the SDO/HMI LOS running difference, the presence of SDO/AIA 1600 \AA\ and 1700 \AA\ wavefronts was determined by eye, in the first instance. An example of such a detection in the SDO/AIA 1600 \AA\ and 1700 \AA\ running difference images can be seen in Fig.~\ref{Fig2}. We initially searched by eye, as all of the SQs we analysed had been detected using the movie method in \cite{Sharykin2020}. Automated methods optimised for detecting SQ signatures in SDO/AIA data could be developed in the future using this sample as a test dataset.

The first step for creating time-distance diagrams from the SDO/AIA data was to find an approximate point-of-origin for each SQ. This was determined for each flare by tracing the SDO/HMI LOS velocity wavefronts back by eye to their apparent source, which coincided with one of the flare foot-points. The flare foot-points were identified as the locations of the brightest emission \cite{Fletcher2004}. Once this epicentre was identified, a pixel range and an angular range in which the wavefront propagated were selected. The angular range was varied to make sure that the clearest SQ ridge would be present in the time-distance diagram. Once these inputs are selected, the same criteria was then used to create co-spatial and co-temporal time-distance diagrams for the AIA observations. The time-distance diagrams were created by averaging across all pixels in the arc between the previously inferred angular range of interest, for each distance from the origin. Filtering was applied when creating these diagrams to reduce the noise. Following this analysis, several SQs appeared as ridges on the time-distance diagrams. The ridges were fitted with a regression trend of $\alpha*x^{0.5}$, which SQs commonly follow \citep{KosovichevZharkova1998} where $`\alpha'$ has units of $s$ $km^{-\frac{1}{2}}$. Figure~\ref{Fig3} displays an example of such a ridge in the SDO/HMI LOS velocity time-distance diagram (A) together with SDO/AIA 1600 \AA\ (B) and SDO/AIA 1700 \AA\ (C) time-distance diagrams. The presence of multiple ridges indicates that multiple, sufficiently strong wavefronts are present (see \citealt{Quinn2019}).

\begin{figure*}
\plotone{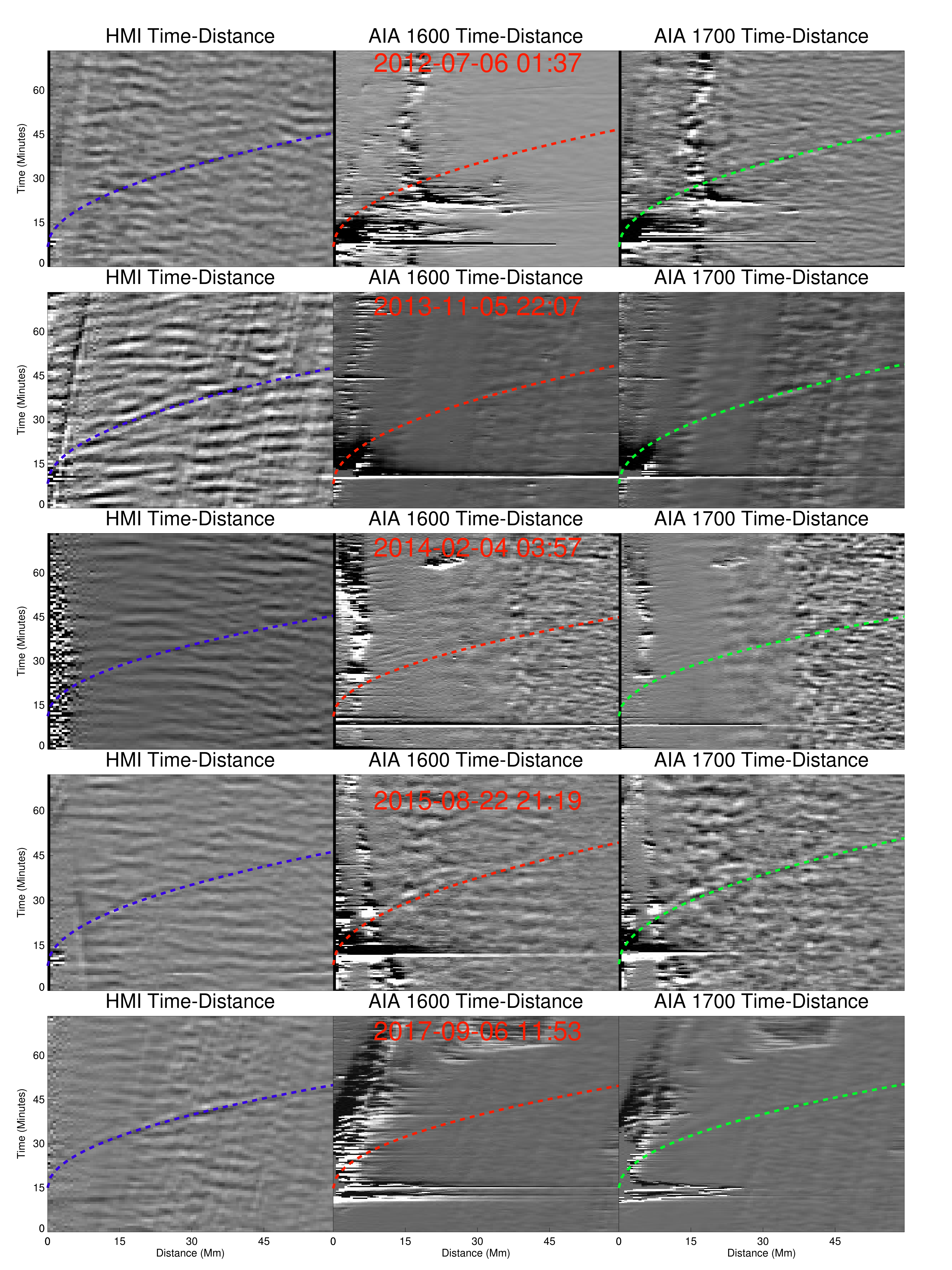}
\caption{The time-distance diagrams created for 5 `Clear' responses (see Table~\ref{Tab1}). The SQs are difficult to detect in the full time-distance diagrams but become more apparent when these diagrams are studied in detail. The approximate locations of the SQs (and the regression trends) are highlighted in blue for the SDO/HMI LOS velocity panels, red for the SDO/AIA 1600 \AA\ panels, and green for SDO/AIA 1700 \AA\ panels to help the reader. It can be seen that the ridge in the SDO/HMI LOS velocity diagrams is co-spatial and co-temporal to the ridges in the SDO/AIA 1600 \AA\ and 1700 \AA\ diagrams. The flares in question, are labelled in red in the middle panel of each row.}
\label{Fig4}
\end{figure*}

\section{Results and Discussion}
\label{sec:Results}

Of the 62 flares that were investigated (see Table~\ref{Tab3} in the Appendix), all had a SQ observable in running-difference SDO/HMI LOS images as expected. Of these events, 25 SQs had some varying degree of visibility by eye in the SDO/AIA 1600 \AA\ and 1700 \AA\ channels. A detailed list of these 25 SQs is provided in Table~\ref{Tab1}. The approximate time at which each SQ is first visible in the filtered SDO/HMI LOS velocity running differences and the approximate direction of propagation of the wavefront, are included for reference. Five of these SQs had very clear wavefronts in running difference images (see middle and bottom row of Fig.~\ref{Fig2} for an example). These responses were approximately co-spatial and co-temporal with the SQs in the SDO/HMI LOS velocity images. These events are labeled `Clear'. Additionally, 12 SQs had comparatively faint responses in the SDO/AIA 1600 \AA\ and 1700 \AA\ running difference images. These are labeled as `Faint'. Finally, eight datasets displayed a faint SDO/AIA 1700 \AA\ response, but no SDO/AIA 1600 \AA\ response and are labeled `Potential' responses. We note that although the majority of these SQ were detectable in unfiltered running difference images, 6 only possessed a possible response once the data had been frequency filtered. These events are marked with a `$\dagger$' in Table~\ref{Tab1}.

In order to better categorise these flare/SQ events, we created time-distance diagrams for the SDO/HMI LOS velocity observations and both SDO/AIA channels for all responses listed in Table~\ref{Tab1}. If the characteristic SQ ridge was present in the SDO/AIA time-distance diagrams for the faint and potential responses, co-spatial and co-temporal to the SDO/HMI LOS velocity time-distance diagrams, we classify that SQ to have had a `Visible' response in the sixth column of Table~\ref{Tab1}. Our analysis revealed that four of the faint and potential responses in running difference images had a SQ ridge present in both the SDO/AIA 1600 \AA\ and 1700 \AA\ time-distance diagrams (these have been marked by an asterisk in Table~\ref{Tab1}), bringing the total number of SQs with obvious responses in the SDO/AIA UV filters to 9. Additionally, four more of the faint and potential responses did have a possible ridge, in one or both AIA diagrams, however, these were deemed to be inconclusive and have not been counted as a `Visible' responses (labelled `Unclear' in Table~\ref{Tab1}). It is worth noting, one flare (2014 February 04 at 03:57 UT) produced its most apparent SQ response by eye in the SDO/AIA images, in a different direction to the most apparent SDO/HMI LOS velocity response, with an angular difference of approximately 45 degrees. These SDO/AIA responses did have a co-temporal and co-spatial SDO/HMI LOS velocity response, but these SQ ripples were not the most obvious of those detectable in the SDO/HMI LOS running difference images for that flare.

A $\chi^2$ test was performed to determine the best regression fits to the SQs detected in the time-distance diagrams for all flare/SQ pairs. The $\chi^2$($=\Sigma((observed-expected)^2/expected)$) test was conducted to measure the difference between a number of pixels that the SQ ridge was present in and the corresponding positions of the expected regression trend ($\alpha*x^{0.5}$). In order to determine the closest regression trend, we calculated $\chi^2$ values across a range of `$\alpha$' values. The lowest $\chi^2$ value was considered to be the best fit, with velocity and acceleration values then being inferred from that regression trend. The red crosses in Fig.~\ref{Fig3} indicate the points along the SQ that were used for the $\chi^2$ test for this event. These points were selected manually, at points corresponding to the strongest SQ ridge in each of the diagrams from the different passbands. They differ slightly between instruments, and are considered the `observed values' for each SQ. The blue regression trends over-laid on Fig.~\ref{Fig3} are considered to be the `expected values', with the fit being the best fit for this event. Nine points were selected for each time-distance diagram, although the spatial size of the fit for each diagram depended on the spread of these points. We only selected points that were unequivocally created by the helioseismic responses of each SQ, and were present in the SDO/HMI LOS velocity time-distance diagram and both SDO/AIA time-distance diagrams, respectively. 

Five further examples of SQs that were deemed to have `Visible' responses in time-distance diagrams are plotted in Fig.~\ref{Fig4}. We note that 2 of the SQs plotted in Fig.~\ref{Fig4} have helioseismic waves visible by eye in running difference images (third and fifth rows), whereas the other 3 SQs were only apparent after analysing the time-distance diagrams. In addition to the SQs highlighted by the regression trends over-laid on Fig.~\ref{Fig4} there are also inclined ridges in the top and bottom AIA diagrams. These are more difficult to see in the full time-distance diagrams but are clearly present upon closer inspection. These inclined ridges resemble the atmospheric, acoustic-gravity waves modelled in \cite{StefanKosovichev2020}. These features could have also been created by the flare ribbon entering into the regions from which the diagrams were created. 

The regression trend of the non-degraded time-distance diagrams derived from the SDO/HMI LOS velocity and SDO/AIA UV filters are different by a factor of 1.875, due to the different cadence of the instruments. As such, we temporally degraded the SDO/AIA 1600 \AA\ and 1700 \AA\ data to match the cadence of the SDO/HMI LOS velocity data. New time-distance diagrams were created for both SDO/AIA 1600 \AA\ and 1700 \AA\ channels. By degrading the SDO/AIA data, we were able to accurately determine whether the ridges were co-spatial and co-temporal with SDO/HMI.  We expected that each pair of SDO/AIA time-distance diagrams would have the same ridge, as these passbands are sampling similar layers of the solar atmosphere. This would suggest that the wavefronts have travelled the same distance and the ridges on their time-distance diagrams should be at the same position. We found that to be the case.

Once again, a $\chi^2$ test was undertaken for each degraded SDO/AIA regression fit. Interestingly, some of the ridges that were clear in the non-degraded time-distance diagrams were no longer visible in the degraded diagrams. We attributed this to the reduction in temporal resolution obscuring the ridges. Table~\ref{Tab2} shows the $\chi^2$ values obtained for the 9 flares, before and after their cadence had been degraded to match the SDO/HMI LOS data. The $\chi^2$ for all visible degraded trends were less than 2, indicating good fits. We note that the SDO/AIA maximum apparent velocities and accelerations were calculated from the temporally degraded data so that the values could be more easily compared to those inferred from the SDO/HMI LOS velocities. Errors were calculated by shifting the `$\alpha$' value by one data-point ($\pm0.249$ s km$^{-\frac{1}{2}}$ in the SDO/HMI fits and $\pm0.182$ s km$^{-\frac{1}{2}}$ for the SDO/AIA data) in each direction from the best fit. No errors are reported for the accelerations as these were below 3 decimal places. Two of these SQs had no signature in the temporally degraded data. No maximum apparent velocities and accelerations, or degraded $\chi^2$ values are reported for these events. We refer to the measured maximum velocities and accelerations as apparent as these wavefronts are not moving horizontally across the surface of the Sun, but are thought to be the reflections from the deeper internal layers of an initial photospheric perturbation \citep{Kosovichev2006}. Overall, we find good agreement in the maximum apparent velocities and accelerations determined from SDO/HMI and SDO/AIA. 

In observations of three of the SQs, the HMI and AIA velocities and accelerations are not in exact agreement with each other, the HMI velocities and accelerations seem to always be higher. This is further highlighted by the slight shift in their regression trends in figure~\ref{Fig4}, with the HMI ridges leaving the FOV slightly earlier than their AIA counterparts. The exact cause for this difference is unclear, and may be related to the different heights that the signal originates from.  

\section{Conclusions}
\label{sec:Conclusions}

We analysed SDO/AIA 1600 \AA\ and 1700 \AA\ and SDO/HMI LOS velocity observations of 62 flares of Solar Cycle 24 and find that nine of these events show clearly detectable SQ signatures. The SDO/AIA UV channels typically sample plasma in the upper photosphere, several hundred km above the region sampled by the SDO/HMI instrument, allowing us to infer SQ properties higher in the atmosphere. The flares analysed here range from low M-class to high X-class but there is no clear correlation between the GOES classification and the detection of the SQ. The apparent lack of correlation between the presence of a SQ and the driving flare GOES class has also been mentioned by \cite{Kosovichev2006}. Time-distance analysis revealed that the maximum apparent transverse velocities in the FOV and accelerations for these 9 responses range between 30.8 {km s$^{-1}$} -- 41.0 {km s$^{-1}$} and 0.011 {km s$^{-2}$} -- 0.020 {km s$^{-2}$}, respectively, for both instruments. However, previous studies have shown that such helioseismic waves can significantly exceed the maximum apparent velocities reported here \citep{Kosovichev2006}.

\cite{Simoes2019} carried out a thorough investigation of the temperature sensitivity of the SDO/AIA instrument and concluded that the flare ribbon emission in the 1600 \AA\ and 1700 \AA\ channels originates from the chromosphere and transition region. It is, therefore, possible that the SQs detected in these channels could be located in the chromosphere, similar to the event discussed by \citet{Quinn2019}. The first SQ response was detected at distances of 5.2 Mm to 25.7 Mm from the flare ribbons for the 5 flares seen in running differences. Their close correspondence with the SDO/HMI LOS observations allow us to conclude that the observed signatures reflect SQ signatures above the photosphere.  

The SQ signatures in the SDO/AIA 1600 \AA\ and AIA 1700 \AA\ images should not be confused with  other atmospheric wave phenomena, such as Moreton waves \citep{Moreton60, Chen2011}. Moreton waves are known to have velocities of up to 500 - 2000 {km s$^{-1}$} and propagate across very large distances (5x10$^5$ km) \citep{Chen2011}. These are distinctly different from our results which have maximum apparent transverse velocities in the FOV of approximately 41 km s$^{-1}$ and propagate over smaller distances (approximately 10 Mm). These oscillations are also distinct from those mentioned in \cite{Milligan2017}, not possessing a 3-minute oscillation, and appearing away from the flare foot-points. As such responses in the SDO/AIA 1600 \AA\ and 1700 \AA\ passbands have not been widely reported, any future work could attempt filtering at frequencies other than 6 mHz to evaluate if  they yield additional detections.

\begin{acknowledgements}
S.Q. acknowledges support from the Northern Ireland Department for the Economy for the award of a PhD studentship. M.M., C.J.N. and A.R. acknowledge support from the UK Science and Technology Facilities Council, under grant No. ST/P000304/1. R.O.M. would like to thank the STFC for the award of the Ernest Rutherford Fellowship (ST/N004981/1). D.B.J. wishes to thank Invest NI and Randox Laboratories Ltd. for the award of a Research $\&$ Development Grant (059RDEN-1), in addition to the STFC for the award of a Consolidated Grant (ST/T00021X/1). We would like to thank the anonymous referee for their comments and suggestions on the manuscript. 

Data supplied courtesy of the SDO/HMI and SDO/AIA consortia. SDO is the first mission to be launched for NASA's Living With a Star (LWS) Program.
\end{acknowledgements}

\appendix

 \begin{longtable}{| c | c | c | c | c |}
\hline
Flare Start Time (UT) & GOES Class & AR & X Location (Arcsecs) & Y Location (Arcsecs)\\

\hline
\hline
$*$ 15.02.2011 01:44 & X2.2 & 11158  & 197 & -224 \\
\hline
18.02.2011 09:55 & M6.6 & 11158  & 749 & -312 \\
\hline
$*$ 30.07.2011 02:04 & M9.3 & 11261  & -497 & 195 \\
\hline
07.09.2011 22:32 & X1.8 & 11283  & 486 & 118 \\
\hline
25.09.2011 08:46 & M3.1 & 11302  & -666 & 137  \\
\hline
26.09.2011 05:06 & M4.0 & 11302  & -558 & 138 \\
\hline
02.10.2011 17:19 & M1.3 & 11302  & 778 & 158 \\
\hline
03.11.2011 20:16 & X1.9 & 11339  & -794 & 281 \\
\hline
09.03.2012 03:22 & M6.3 & 11429  & -5 & 409 \\
\hline
09.05.2012 21:01 & M4.1 & 11476  & -417 & 250 \\
\hline
10.05.2012 04:11 & M5.7 & 11476  & -361 & 260 \\
\hline
04.07.2012 09:47 & M5.3 & 11515  & 284 & -288 \\
\hline
04.07.2012 12:07 & M2.3 & 11515  & 332 & -313 \\
\hline
$*$ 04.07.2012 14:35 & M1.3 & 11515  & 334 & -313 \\
\hline
05.07.2012 01:05 & M2.4 & 11515  & 392 & -313 \\
\hline
05.07.2012 03:25 & M4.7 & 11515  & 436 & -312 \\
\hline
05.07.2012 06:49 & M1.1 & 11515  & 392 & -313 \\
\hline
05.07.2012 10:44 & M1.8 & 11515  & 512 & -313 \\
\hline
05.07.2012 11:39 & M6.1 & 11515  & 496 & -313 \\
\hline
05.07.2012 20:09 & M1.6 & 11515  & 512 & -313 \\
\hline
$*$ 06.07.2012 01:37 & M2.9 & 11515  & 576 & -323 \\
\hline
06.07.2012 13:26 & M1.2 & 11515  & 632 & -313 \\
\hline
06.07.2012 23:01 & X1.1 & 11515  & 812 & -253 \\
\hline
09.07.2012 23:03 & M1.1 & 11520  & -528 & -328 \\
\hline
$*$ 23:10:2012 03:13 & X1.8 & 11598  & -777 & -265 \\
\hline
17.02.2013 15:45 & M1.9 & 11675  & -343 & 305 \\
\hline
24.10.2013 10:30 & M3.5 & 11875  & 349 & 31 \\
\hline
05.11.2013 18:08 & M1.0 & 11890  & -639 & -231 \\
\hline
$*$ 05.11.2013 22:07 & X3.3 & 11890  & -639 & -231 \\
\hline
06.11.2013 13:39 & M3.8 & 11890  & -533 & -276 \\
\hline
07.11.2013 03:34 & M2.3 & 11890  & -444 & -283 \\
\hline
07.11.2013 14:15 & M2.4 & 11890  & -342 & -283 \\
\hline
08.11.2013 04:20 & X1.1 & 11890  & -214 & -275 \\
\hline
10.11.2013 05:08 & X1.1 & 11890  & 238 & -266 \\
\hline
07.01.2014 10:07 & M7.2 & 11944  & -361 & -269 \\
\hline
02.02.2014 06:24 & M2.6 & 11968  & -297 & 314 \\
\hline
$*$ 04.02.2014 03:57 & M5.2 & 11967  & 160 & -98 \\
\hline
07.02.2014 10:25 & M1.9 & 11968  & 695 & 292 \\
\hline
16.02.2014 09:20 & M1.1 & 11977  & -6 & -56 \\
\hline
08.05.2014 09:59 & M5.2 & 12056  & -737 & 112 \\
\hline
20.10.2014 18:55 & M1.4 & 12192  & -654 & -478 \\
\hline
22.10.2014 01:16 & M8.7 & 12192  & -344 & -268 \\
\hline
22.10.2014 14:02 & X1.6 & 12192  & -224 & -268 \\
\hline
26.10.2014 18:07 & M4.2 & 12192  & 599 & -292 \\
\hline
20.12.2014 00:11 & X1.8 & 12242  & 471 & -277 \\
\hline
03.01.2015 09:40 & M1.1 & 12253  & -302 & -41 \\
\hline
10.03.2015 03:19 & M5.1 & 12297  & -615 & -195 \\
\hline
15.03.2015 09:36 & M1.0 & 12297  & 404 & -205 \\
\hline
25.06.2015 08:02 & M7.9 & 12371  & 487 & 67 \\
\hline
$*$ 22.08.2015 21:19 & M3.5 & 12403  & -229 & -349 \\
\hline
28.09.2015 14:53 & M7.6 & 12422  & 406 & -446 \\
\hline
29.09.2015 06:39 & M1.4 & 12422  & 542 & -420 \\
\hline
30.09.2015 13:18 & M1.1 & 12422  & 702 & -389 \\
\hline
04.09.2017 20:28 & M5.5 & 12673  & 216 & -260 \\
\hline
05.09.2017 01:03 & M4.2 & 12673  & 252 & -260 \\
\hline
06.09.2017 08:57 & X2.2 & 12673  & 527 & -247 \\
\hline
$*$ 06.09.2017 11:53 & X9.3 & 12673  & 557 & -265 \\
\hline
07.09.2017 04:59 & M2.4 & 12673  & 677 & -235 \\
\hline
07.09.2017 10:11 & M7.3 & 12673  & 677 & -235 \\
\hline
07.09.2017 14:20 & X1.3 & 12673  & 732 & -232 \\
\hline
08.09.2017 02:19 & M1.3 & 12673  & 799 & -231 \\
\hline
08.09.2017 07:40 & M8.1 & 12673  & 759 & -215 \\
\hline
\caption{The $62$ flares that matched our selection criteria and were, therefore, studied in this article. Running difference images constructed from the SDO/AIA 1600 \AA\ and 1700 \AA\ filters were analysed for each flare in order to investigate whether any signature of SQs was present. This list includes all flares that produced a SQ signature in SDO/HMI running difference images from Table 1 in \cite{Sharykin2020}. Those flares marked with a star were flares that produced a `Visible' (see Table~\ref{Tab1}) solar atmosphere response.}
\label{Tab3}
\end{longtable}

\end{document}